\documentclass[12pt]{article} \input epsf.tex
\usepackage[dvipdfmx]{graphicx} \usepackage{bmpsize}
\usepackage{amsmath,amsfonts,amssymb,stmaryrd,latexsym}
\usepackage{url}
\renewcommand{\theequation}{\thesection.\arabic{equation}}
\newcommand{\be}{\begin{equation}}
\newcommand{\ee}{\end{equation}}
\newcommand{\bea}{\begin{eqnarray}}
\newcommand{\eea}{\end{eqnarray}}
\newcommand{\bear}{\begin{eqnarray}}
\newcommand{\eear}{\end{eqnarray}}
\newcommand{\met}{E_T \!\!\!\!\!\! \slash  \,\,\, }
\newcommand{\ba}{\begin{array}}
\newcommand{\ea}{\end{array}}

\def\tev{\,{\rm TeV}}

\def\SU2{\rm SU(2)}
\def\U1{\rm U(1)}

\newcommand{\ie}{{\it i.e.}}

\newcommand{\gR}{g_R}
\usepackage[usenames,dvipsnames]{xcolor}

\begin{document}

\baselineskip=18pt \pagestyle{plain} \setcounter{page}{1}

\vspace*{-1cm}

\noindent \makebox[11.5cm][l]{\small \hspace*{-.2cm} }{\small Fermilab-Pub-15-430-T}  \\  [-1mm]

\begin{center}

{\Large \bf  Signals of a 2 TeV $W'$ boson and a heavier $Z'$ boson   } \\ [9mm]

{\normalsize \bf Bogdan A. Dobrescu and Patrick J. Fox \\ [3mm]
{\small {\it
Theoretical Physics Department, Fermilab, Batavia, IL 60510, USA   
  }}\\
}

\center{November 5, 2015; revised April 8, 2016 }

\end{center}

\vspace*{0.2cm}

\begin{abstract}

We construct an $SU(2)_L\times SU(2)_R\times U(1)_{B-L}$  
model with a Higgs sector that 
consists of a bidoublet and a doublet, and with
a right-handed neutrino sector that includes  one Dirac fermion and one Majorana fermion.
This model explains the Run 1 CMS and ATLAS excess events in the $e^+e^-jj$, $jj$, $Wh^0$ and $WZ$ channels in terms of a 
$W'$ boson of mass near 1.9 TeV and of coupling $g_R$ in the 0.4--0.5 range, with the lower half preferred by limits on $t \bar b$ resonances and Run 2 results.
The production cross section of this $W'$  boson at the 13 TeV LHC is in the 700--900 fb range, allowing sensitivity in more than 17 final states.
We determine that the $Z'$ boson has a mass in the 3.4--4.5 TeV range and several decay channels that can be probed in Run 2 of the LHC, including 
cascade decays via heavy Higgs bosons.

\end{abstract}



{\small
\tableofcontents
}

\newpage

\section{Introduction} \setcounter{equation}{0}

The field content of the Standard Model (SM) of particle physics includes fermions, gauge bosons and a Higgs field. Quantum field theories whose low-energy 
limit is given by the SM often include a larger gauge group, an extended Higgs sector and additional fermions. Various theoretical constraints link
the properties of these new fields, especially in connection to gauge symmetries and their spontaneous breaking.
For example, the discovery of a $W'$ boson (spin-1 field of electric charge $\pm 1$) would imply the existence of a $Z'$ boson (electrically-neutral spin-1 field), additional Higgs particles, and in many cases additional fermions.

Several excess events reported by the CMS  \cite{Khachatryan:2014dka,CMS:2015gla,Khachatryan:2015sja,Khachatryan:2014gha} 
and ATLAS  \cite{Aad:2015owa}  Collaborations based on the LHC Run 1 data hint towards a $W'$ boson of mass in the 1.8--2 TeV range.
An explanation for all these excess events is presented in \cite{Dobrescu:2015qna} based on the $SU(2)_L \times SU(2)_R \times U(1)_{B-L}$ gauge 
group \cite{Marshak:1979fm}, with a Higgs sector consisting of a bidoublet scalar and an $SU(2)_R$-triplet scalar.
The phenomenological implications of that Higgs sector are discussed in \cite{Dobrescu:2015yba}, where it is shown that the $W'$ decays 
into heavy Higgs bosons provide an explanation for a peculiar excess reported by ATLAS in final states with two leptons of the same 
charge and two or more $b$ jets \cite{Aad:2015gdg}.

The $SU(2)_L \times SU(2)_R \times U(1)_{B-L}$ symmetry implies the existence of ``right-handed" neutrinos. These usually  have Majorana masses, 
or else have Dirac masses equal to those of the SM neutrinos. The CMS excess in the 
$e^+e^-jj$ final state, and the absence of an $e^+e^+ jj$ excess imply that at least one of the right-handed neutrinos has a mostly Dirac mass at the TeV scale.
As the right-handed neutrinos are two-component fermions, the question is which fermion fields do they partner with to become Dirac fermions? 
In  \cite{Dobrescu:2015qna,Dobrescu:2015yba,Dev:2015pga} 
it is proposed that some new fermions become the Dirac partners of the  right-handed neutrinos. A more intriguing possibility 
is that the electron and tau right-handed neutrinos form together a Dirac fermion as a result of a lepton-flavor symmetry \cite{Coloma:2015una}; this is related 
to the observation \cite{Gluza:2015goa} that there is a special point in the parameter space of  right-handed neutrino Majorana masses where the $e^+e^+ jj$ signal is 
suppressed.

Here we present a simpler $SU(2)_L \times SU(2)_R \times U(1)_{B-L}$ model that explains the CMS and ATLAS excess events. 
The Higgs sector includes only a bidoublet and an $SU(2)_R$-doublet,
which allows the $Z'$ boson to have a mass as low as the current limit, of roughly 3 TeV. 
The right-handed neutrino sector includes a global symmetry forcing
the electron and tau right-handed neutrinos to form a Dirac fermion. 
We analyze the LHC signatures of this model, using Monte-Carlo simulations, and extract the region of parameter space consistent with the observed deviations from the SM.
We find that  simulations of different $W'$ decay channels are required for this extraction due to contamination among signals.

In Section 2 we describe the model, we extract the value of the $W'$ coupling consistent with the CMS dijet excess \cite{CMS:2015gla}, and we compute the $W'$ 
width and production cross section. 
In Section 3 we discuss the right-handed neutrino sector, and we show that the CMS $e^+e^- jj$ excess is accommodated for a range of right-handed neutrino masses.
$W'$ decays into pairs of bosons are analyzed in Section 4. The properties of the $Z'$ boson are studied in Section 5. Our conclusions 
are summarized in Section 6.

\section{$W'$ width and production cross section} 
\label{sec:Wpwidthandproduction}

We consider an $SU(3)_c \times SU(2)_L\times SU(2)_R\times U(1)_{B-L}$ gauge theory with the field content
shown in Table \ref{table:fields}. 
The right-handed neutrinos $N_R^i$ ($i$ labels the three generations) acquire masses at the TeV scale through the mechanism of Ref.~\cite{Coloma:2015una}, 
also discussed in section 3. The Higgs sector includes an $SU(2)_R$ doublet $H_R$ whose VEV, of several TeV, 
breaks $SU(2)_R\times U(1)_{B-L}$ down to the SM hypercharge group $U(1)_Y$, and a bidoublet $\Sigma$ whose 
VEV breaks $SU(2)_L\times U(1)_{Y}$ at the weak scale.

\begin{table}[b!]
\renewcommand{\arraystretch}{1.7}
\vspace*{2mm}
\centering
\begin{tabular}{|c|c||c|c|c|c|}
\hline
Fields & spin & $SU(3)_c$ & $SU(2)_L$ & $SU(2)_R$ & $U(1)_{B-L}$ \\ \hline\hline
$q_L^i=(u_L^i,~d_L^i)^\top$ & 1/2 &  3 &  2 & 1 & +1/3 \\ 
$q_R^i=(u_R^i,~d_R^i)^\top$ & 1/2 & 3 &   1 & 2 & $+1/3$ \\ \hline
$L_L^i=(\nu_L^i,~\ell_L^i)^\top$ &  1/2 & 1 & 2 & 1 & $-1$ \\ \
$L_R^i=(N_R^i,~\ell_R^i)^\top$ &  1/2 & 1 & 1 & 2 & $-1$ \\  \hline \hline
$\Sigma$ &  0 & 1 & 2 & 2 & 0 \\ 
$H_R$ & 0 & 1 & 1 & 2 & 1 \\ \hline 
\end{tabular}
\caption{\small Fields carrying $SU(3)_c \times SU(2)_L\times SU(2)_R\times U(1)_{B-L}$ gauge charges. The index $i=1,2,3$ labels the three generations of fermions. }
\label{table:fields}
\end{table}

The only gauge fields beyond the SM are a $W'$ boson and a $Z'$ boson, 
which acquire masses primarily due to the $H_R$  VEV. For $M_{W'} \approx 1.9$ TeV, as indicated by the 8 TeV LHC data,  the $Z'$ boson has a mass of a few TeV
 (discussed in Section 5).
The VEV of the bidoublet $\Sigma$ leads to mass mixing between the  $SU(2)_L\times SU(2)_R$ gauge bosons.
The $W$ boson and the hypothetical 
$W'$ boson are linear combinations of the electrically-charged $SU(2)_L\times SU(2)_R$ gauge bosons, $W_L^{\pm \mu}$ and $W_R^{\pm \mu}$,
\be
\left( \begin{array}{c}    W^+_\mu  \\ [4mm] W^{\prime +}_\mu  \end{array} \right) =
 \left( \begin{array}{cc}  \cos\theta_{\!_+}      \;  &  \;    \sin\theta_{\!_+}    \\ [5mm]
      	                          -  \sin\theta_{\!_+}             \; &  \;   \cos\theta_{\!_+}    
    \end{array} \right)   
    \left( \begin{array}{c}    W_{L\, \mu}^+  \\ [4mm]  W_{R\, \mu}^+  \end{array} \right)
    ~~,
\label{eq:wp}
\ee
with the $W_L-W_R$ mixing angle $\theta_{\!_+}$ given by  \cite{Dobrescu:2015yba}
\be
\theta_{\!_+} = \frac{g_R}{g } \left( \frac{ M_W}{ M_{W'}}\right)^{\! 2} \sin 2\beta  \left(1 + O(M_W^2/M_{W'}^2) \right) ~~.
\ee
The parameters introduced here are as follows: $g_R$  is the $SU(2)_R$ gauge coupling, $g\approx 0.63$ is the SM weak coupling at 2 TeV,  and 
the angle $\beta$  arises from the usual $\tan\beta$ ratio of VEVs in the Two-Higgs doublet model, which is the manifestation of 
the bidoublet scalar $\Sigma$  below the $SU(2)_R$ breaking scale. 

The couplings of the $W'$ boson to pairs of SM fermions, neglecting the $W_L -W_R$ mixing,  are given by 
\be
\frac{g_R}{\sqrt{2}} \,  W'_\mu  \left( \bar u_R \gamma^\mu d_R + \bar c_R \gamma^\mu s_R + \bar t_R \gamma^\mu b_R \right) + {\rm H.c.}
\label{eq:quarks}
\ee
The quark fields here are in the mass eigenstate basis. Flavor-dependent perturbations to the above interactions are allowed by 
the $SU(2)_R$ gauge symmetry, but are constrained by measurements of flavor processes such as $K-\bar K$ meson mixing,
and we assume here that they are negligible.

The couplings of the $W'$ boson to pairs of SM bosons are fixed by gauge invariance  and take the form  \cite{Dobrescu:2015yba}:
\bear
&& \hspace*{-0.81cm}  {\cal L}_{W'WZ} =\frac{i g_R }{ c_W} \left( \!\frac{M_W}{M_{W'}}\! \right)^{\! 2} \sin2\beta \,   \left[ W^{\prime + }_\mu \! \left(W^-_\nu \partial^{[\nu}Z^{\mu]} +  Z_\nu \partial^{[\mu} W^{-\nu]}  \rule{0pt}{12pt}  \right)
+Z_\nu W^-_\mu \partial^{[\nu} W^{\prime +\mu]}   \rule{0pt}{14pt} \right]  +  {\rm H.c.} ,
\nonumber \\ [3mm]
&& \hspace*{-0.81cm}  {\cal L}_{W'Wh} = - g_R M_W   \sin (2\beta-\delta)  \; W^{\prime +}_\mu \, W^{\mu -}  h^0   +  {\rm H.c.} ,
\label{eq:bosons}
\eear
where $c_W \equiv \cos\theta_W  \approx 0.87$ ($\theta_W$ is the weak mixing angle at the $M_{W'}$ scale), 
and $[\mu,~\nu]$ represents the commutation of Lorentz indices.
The $WZ$ and $Wh$ signals (see Section 4) indicate $0.36 \lesssim \tan\beta \lesssim 2.8 $.  
The agreement between the SM predictions and the measured properties
of $h^0$ requires the mixing angle between the neutral CP-even Higgs bosons, $\alpha$, to be near the alignment limit.  
The amount  by which the mixing deviates from the alignment limit is $\delta = \beta-\alpha-\pi/2 \ll 1$.

The $W'$ partial widths into SM particles induced by the couplings shown in Eqs.~(\ref{eq:quarks}) and  (\ref{eq:bosons}) are given by \cite{Dobrescu:2015yba}
\bear
&& \Gamma(W^\prime \to t \bar b \, ) \simeq \Gamma(W^\prime \to c \bar s \, ) =\Gamma(W^\prime \to u \bar d \, ) 
= \frac {g_R^2} {16\pi} \, M_{W^\prime}  \left( 1 + \frac{\alpha_s}{\pi} \right)  ~~,
\nonumber \\ [3mm]
&& \Gamma(W^\prime \to WZ) \simeq  \Gamma(W^\prime \to Wh^0)  \simeq \frac {g_R^2 } {192\pi}   \sin^2 \! 2\beta \; M_{W^\prime}  ~~.
\label{eq:WZWh}
\eear
The phase-space suppression factors (ignored here) give corrections of order $m_t^2/M_{W'}^2 \approx 0.8\%$ or $(M_h+M_W)^2/M_{W'}^2 \approx 1.2\%$ for $M_{W'} = 1.9$ TeV.
The next-to-leading order (NLO) QCD corrections (included here) are of order 3\%, with $\alpha_s \approx 0.1$ being the QCD coupling constant 
at the $M_{W'}$ scale.

Besides the branching fractions into SM particles, the $W'$ boson can decay into 
a right-handed neutrino and an electron or $\tau$,
and also into two heavy Higgs bosons or one heavy Higgs boson and one SM boson.
The absence of $\mu\mu jj$ signals in Run 1 of the LHC \cite{Khachatryan:2014dka,Aad:2015xaa} implies that the second-generation right-handed neutrino is heavier than (or nearly degenerate to) the $W'$.

The increase of the total width due to decays involving right-handed neutrinos (see Section 3) is between 4\% and 12\% 
(depending on the masses of right-handed neutrinos);  
an additional increase of up to 6\% is due to decays involving heavy Higgs scalars (discussed in Section 4).
Including these contributions, and taking the range for $\tan\beta$ 
into account, the total $W'$ width is
\be
\Gamma_{W'} \approx (31-35) \, {\rm GeV} \; \left(\frac{g_R}{0.5}\right)^{\! 2} \left(\frac{M_{W^\prime}}{1.9 \; {\rm TeV}}\right)  ~~.
\label{eq:GammaWp}
\ee

The $W'$ coupling $g_R$ can be determined from the cross section required to produce the dijet resonance near $M_{W'}$.
The  CMS dijet excess \cite{Khachatryan:2015sja} 
at a mass in the 1.8--1.9 TeV range indicates that the $W'$ production cross section times the dijet branching fraction 
is in the 100--200 fb range (this is consistent with the ATLAS dijet result \cite{Aad:2014aqa}, which shows a smaller excess at 1.9 TeV)\footnote{The  CMS \cite{Khachatryan:2015dcf} dijet search with 2.4 fb$^{-1}$ of Run 2 data has a small excess (less than 2 $\sigma$) in the 1.7--1.8 TeV range that may be consistent with the Run 1 excess. The ATLAS \cite{ATLAS:2015nsi} result with 3.6 fb$^{-1}$ of Run 2 data has no excess in the 1.8--2 TeV range, and sets a 95\% CL upper limit on the dijet production cross section of around 300 fb at $M_{W'} = 1.9$ TeV. Reinterpreting this limit in our model we find
 $g_R \lesssim 0.44$.}. 
This was assumed  in Refs.~\cite{Dobrescu:2015qna,Brehmer:2015cia}
to be the range for $\sigma_{8 jj}(W')  \equiv \sigma ( pp \to W' \! \to jj)_{\sqrt{s}=8 \; {\rm TeV}}$, 
where $j$ is a hadronic jet associated with quarks or antiquarks other than the top. We point out here that 
the large $W'$ mass implies that the CMS dijet search is also sensitive to merged jets arising from the decays of boosted 
top quarks and boosted $W$, $Z$ or $h^0$ bosons produced in $W'$ decays. 

To see that, note that the CMS search \cite{Khachatryan:2015sja} identifies a narrow 
jet of cone $\Delta R \leq 0.5$, and then includes all hadronic activity in a wider cone of  $\Delta R \leq 1.1$. This procedure 
is designed to collect all the radiation associated with the jet so that the invariant mass distribution is properly calibrated.
An additional  consequence of this procedure is that it cannot separate 
$W'$ decays into two light jets from those into heavy SM particles that decay hadronically.

Consider first the $W' \!\to t \bar b$ channel. The case where the top quark decays hadronically 
implies that there is a narrow $b$ jet back-to-back to a merged jet formed of three jets. As the CMS search is not using $b$-tagging,
and the merged jet is typically much narrower than  $\Delta R = 1.1$, this topology will contribute to the dijet peak near $M_{W'}$.

The case of semileptonic top decays is more complicated. The lepton 
is almost collinear with the jet arising from the top decay so that it usually does not pass the isolation requirement.
The missing transverse energy carried by the neutrino is typically large enough to move the invariant mass of the 
system formed by the $\bar b$ jet and the lepton-plus-$b$ jet to values significantly lower than $M_{W'}$.
Thus, the fraction of $W' \to t \bar b$ decays that contribute to the dijet peak near $M_{W'}$ is expected to be roughly given by $B(W\to jj) \approx 67.4\%$.
This would imply that combining the $W' \to t \bar b$ and $W' \to jj$ processes increases the number of dijet events by $\sim$ 34\% 
compared to the $W'\to jj$ process.

To improve this estimate, we have implemented the interactions (\ref{eq:quarks}) and (\ref{eq:bosons}) 
in FeynRules \cite{Alloul:2013bka},  which automatically generated the MadGraph \cite{Alwall:2014hca} model files available at \cite{ourMG5}.
Using the Delphes \cite{deFavereau:2013fsa} detector simulation and the NN23LO1 parton distributions \cite{Ball:2012cx}, we find that 
$W' \to t \bar b$ increases the number of dijet events that pass the selection cuts and have an invariant mass above 1.7 TeV by 37\% compared to $W'\to jj$ by itself.

A similar discussion applies to $W' \to WZ$ or $Wh^0$. Events in which both the $W$ and  $Z$ decay hadronically 
contribute to the dijet peak. These represent $B(W\to jj) B(Z\to jj)  \approx 47.1\%$ of all  $W' \to WZ$ events.
In the case of  $W' \to Wh^0$, the purely hadronic decays lead to a $W$ merged jet back-to-back to a Higgs merged jet.   
The fraction of $W' \to Wh^0$ events that contribute to the dijet peak near $M_{W'}$ is expected to be
$B(W\to jj) B(h^0 \to jj , 4j)  \approx  54.1\%$, where $B(h^0 \to jj , 4j) \approx 80.2\%$ is the total hadronic branching fraction of
the SM Higgs boson for $M_h = 125$ GeV \cite{Heinemeyer:2013tqa}.  Note that decays involving taus include missing energy so that 
their contribution to the dijet peak near $M_{W'}$ is suppressed.

Our simulations show that the contributions from $W' \to WZ$ and $Wh^0$ to the dijet peak are larger than the above 
estimates. However, given the small branching fractions of these modes, the increase in the 
number of dijet events that pass the selection cuts and have an invariant mass above 1.7 TeV is only 4\%.

Finally, the $W'$ decays to heavy Higgs bosons (if kinematically open) also contribute to the dijet signal. The main channels, $W' \to H^\pm A^0$
and $H^\pm H^0$, followed by $H^+ \to t \bar b$ and $A^0/H^0 \to t\bar t$ give two merged jets: one containing $Wb\bar b$ 
and the other one containing $W^+W^-b \bar b$. When these three $W$ bosons decay hadronically (a 31\% combined branching fraction), 
the reconstructed mass of the two merged jets is again near $M_{W'}$. 

Summing over all these modes, we find that the total branching fraction of decays contributing to the dijet peak near $M_{W'}$ is  
$B(W' \to {\rm dijet} ) \approx  81-89\%$,
with the uncertainty arising mainly due to the right-handed neutrino masses. 
The total production cross section at $\sqrt{s} = 8$ TeV required to explain the dijet peak is
\be
\sigma_{8}(W') \approx \frac{100-200 \; {\rm fb}}{B(W' \to {\rm dijet} ) } \approx 112-246 \; {\rm fb}   ~~.
\label{eq:observed}
\ee
This observed range needs to be compared with the predicted cross section. 
For $M_{W'}=1.9$ TeV, the leading-order $W'$ production cross section at $\sqrt{s} =8$ TeV, computed with MadGraph,
is $155 \,  {\rm fb}\;  ( g_R/0.5)^2$. Next-to-leading order QCD effects can be taken into account by multyplying 
the leading-order cross section by $K_8 (W') \approx 1.15$ \cite{Duffty:2012rf} (this is consistent with the result of an MCFM \cite{Campbell:2015qma} computation, but somewhat smaller than the $K$-factor
computed in \cite{Cao:2012ng}).
Thus, the predicted cross section is
\be
\sigma_{8}(W') \approx  178 \,  {\rm fb}\;  \left(\frac{g_R}{0.5}\right)^2  ~~.
\label{eq:predict}
\ee
Comparing Eqs.~(\ref{eq:observed}) and (\ref{eq:predict}) allows us to determine the range for the $W'$ coupling:
\be
0.40 \lesssim  g_R \lesssim 0.59  ~~.
\label{eq:gR}
\ee
Combining this with Eq.~(\ref{eq:GammaWp}) implies a total width $\Gamma_{W'} = 20$--48 GeV.  Note that combining Run 1 and Run 2 dijet results reduces the required production cross section, but there is a theoretical lower bound $g_R>g_Y\approx 0.363$ (see section~\ref{sec:Zpproperties}).

Searches for $tb$ resonances using LHC Run 1 data  impose a stronger upper limit on $g_R$.
The semileptonic top decay is used by CMS  \cite{Chatrchyan:2014koa}
to set a cross section limit $\sigma(pp \to W' \to tb) < 100$ fb at the 95\% CL.
The hadronic  top decay leads to a stronger CMS limit \cite{Khachatryan:2015edz}: $\sigma(pp \to W' \to tb) < 60$ fb at the 95\% CL.
The combination \cite{Khachatryan:2015edz} of these two $tb$ final states gives $\sigma(pp \to W' \to tb) \lesssim 40$ fb for $M_{W'} = 1.9$ TeV. 
This implies
$g_R \lesssim 0.45 $.
The ATLAS searches using semileptonic  \cite{Aad:2014xea} and hadronic \cite{Aad:2014xra} top decays impose weaker
upper limits on $\sigma(pp \to W' \!\to tb)$, of 120 fb and 210 fb, respectively. The discrepancy between the CMS and ATLAS 
$tb$ sensitivities in the case of hadronic top decays is due in part to the $b$ tagging efficiencies, which are not well 
understood in the case of the large jet $p_T$  associated with a resonance near 2 TeV. 

The $W'$ production at the 13 TeV LHC is 6.3 times larger than in Run 1, 
\be
\sigma_{13}(W') \approx  1130 \,  {\rm fb}\;  \left(\frac{g_R}{0.5}\right)^2  ~~,
\label{eq:predict13}
\ee
taking into account a $K$-factor of $K_{13}(W') \approx 1.2$ \cite{Sullivan:2013ina}.  For $0.4<g_R<0.44$ the Run 2 production cross section lies between 720 and 875 fb. 
Consequently, the existence of the $W'$ can be tested in the $W' \!\to jj$ and $W' \!\to t b$ channels with $O(10)$ fb in Run 2.

\medskip

\section{$W' \to e N_D$ and $N_D$ decay widths}
\label{sec:eN} \setcounter{equation}{0}

An important component of a complete $SU(2)_L \times SU(2)_R \times U(1)_{B-L}$ model is the sector responsible for the right-handed neutrino masses.
In this Section we analyze a mechanism that allows the three right-handed neutrinos to acquire masses 
consistent with the CMS $e^+e^-jj$ excess \cite{Khachatryan:2014dka} and the lack of a signal in related LHC searches (especially in the same-sign 
$eejj$ final state \cite{Keung:1983uu}). This mechanism does not require any field 
of mass below the $SU(2)_R \times U(1)_{B-L}$ breaking scale ($u_H \sim$ 5--7 TeV)  beyond those listed in Table 1. This relies, though, 
on higher-dimensional operators whose origin requires additional fields (discussed at the end of this Section) 
that have masses below $\sim  20$ TeV.

Let us consider the following lepton-number-violating dimension-5 operators:
\be   
- \frac{ C_{e\tau} }{m_\psi^{e\tau}} (\overline L^e_R)^c H_R   \widetilde H_R L^\tau_R -
\frac{C_{\mu\mu} }{2 m_\psi^{\mu\mu} }  (\overline L^\mu_R)^c H_R  \widetilde H_R L^\mu_R 
+ {\rm H.c.}  ~~,
\label{eq:ops}
\ee
where $ \widetilde H_R \equiv i \sigma_2 H_R^*$. 
The flavor structure of these operators is a consequence of the flavor symmetry introduced in \cite{Coloma:2015una}: 
a global $U(1)$ symmetry with $L^e_R, L^\mu_R, L^\tau_R$ carrying charge $-1, 0, +1$, respectively. All other fields shown in Table 1 have global $U(1)$ charge 0.
Replacing the $SU(2)_R $ doublet $H_R$ by its VEV in Eq.~(\ref{eq:ops}) leads to the following mass matrix for right-handed neutrinos:
\be
- u_H^2 \left[  \frac{ C_{e\tau} }{m_\psi^{e\tau}}  \!\left( (\overline N^e_R)^c  N^\tau_R +  (\overline N^\tau_R)^c  N^e_R \rule{0mm}{5mm}\right) 
+ \frac{C_{\mu\mu} }{m_\psi^{\mu\mu} }  (\overline N^\mu_R)^c  N^\mu_R \right]  ~~.
\label{eq:massterms}
\ee
Although these are Majorana masses, the $N^e_R$ and $N^\tau_R$ fields form a Dirac fermion \cite{Coloma:2015una}, labelled here $N_D$.
This has a mass $m_{N_D} = C_{e\tau} u_H^2 / m_\psi^{e\tau}$, and its left- and right-handed components are given by
$N_{D_L} \equiv  (\overline N^\tau_R)^c$ and $N_{D_R} \equiv  N^e_R$, respectively. The muon right-handed neutrino $N^\mu_R$
remains a Majorana fermion of mass $m_{N^\mu} = C_{\mu\mu} u_H^2 / m_\psi^{\mu\mu}$.

The couplings of the $W'$ boson to the Dirac fermion $N_D$, or the muon right-handed neutrino 
$N^\mu_R$,  and the SM charged leptons  in the mass eigenstate basis follow from
 the $SU(2)_R$ gauge symmetry:
\be
\frac{g_R }{\sqrt{2}} \,  W'_\nu  \;  \left[ \overline N_{D_R} \gamma^\nu \, e_R
+ (\overline N_{D_L} )^c \gamma^\nu \, \tau_R + \overline N_R^\mu \gamma^\nu \, \mu_R \rule{0mm}{5mm} \right]
+ {\rm H.c.} ~~,
\label{eq:leptons}
\ee
where the $W'$ coupling $g_R$ is the same as in Eq.~(\ref{eq:quarks}).

As no $\mu\mu jj$ signal has been observed yet,  we will assume the following mass hierarchy:
$m_{N_D} < M_{W'} \lesssim m_{N^\mu}$.  As a result, the $W' \! \to \mu \, N^\mu$ decay is kinematically closed or strongly phase-space suppressed, 
and we will ignore the presence of $N^\mu_R$  in what follows.
The $W'$ partial widths into a SM lepton and an $N_D$ Dirac fermion follow from Eq.~(\ref{eq:leptons}):
\be
\Gamma \left( W' \! \to  e N_D \right) = \Gamma \left( W' \! \to \tau N_D \right) = 
 \frac {g_R^2} {48\pi} \, M_{W^\prime}  \left( 1 + \frac{m_{N_D}^2}{2 M_{W'}^2} \right)  \left( 1 - \frac{m_{N_D}^2}{M_{W'}^2} \right)^{\! 2}  ~~.
\ee

The $W'$ couplings shown in Eq.~(\ref{eq:leptons}) also induce 3-body decays of the Dirac fermion $N_D$, through an off-shell $W'$. These 
have equal widths into $e^- u \bar d$ and $e^- c \bar s$ given by\footnote{Our result to leading order in $m_{N_D}$ 
is larger by a factor of 2 than the corresponding one implied by Eq.~(13) of \cite{delAguila:2009bb}, after 
taking into account the difference between  Majorana and Dirac fermions (assuming that the $l jj$ final state is defined there as the sum over $l = e,\mu,\tau$).}
\be  \hspace*{-3mm}
\Gamma \left( N_D \to e^- W^{\prime *}\! \to e^- u \, \bar d  \, \right) = \frac{g_R^4  m_{N_D}^5  }{ 2048\pi^3 M_{W'}^4 \! } 
\left( 1 + \frac{3 m_{N_D}^2 }{5 M_{W'}^2} + \frac{2 m_{N_D}^4 }{5 M_{W'}^4} + O(m_{N_D}^6 / M_{W'}^6) \right)  .
\label{eq:3body}
\ee
For $m_{N_D}$ close to $M_{W'}$ one has to use the exact formula for this width given in the Appendix.
The width for $N_D \to e^- t \bar b $ is smaller:
\be 
\Gamma \left( N_D \to e^- W^{\prime *} \!\to e^- t \, \bar b  \, \right)  \simeq
\Gamma \left( N_D \to e^- W^{\prime *} \!\to e^-  u \, \bar d  \, \right) f(m_{N_D}) 
~~,
\label{eq:3bodytb}
\ee
where the phase-space suppression is included in 
\be
f(m_{N_D}) = 1 - \frac{8 m_t^2 }{ m_{N_D}^2} \left( 1 - \frac{9 m_{N_D}^2 }{40 M_{W'}^2}  + O(m_{N_D}^4 / M_{W'}^4) \right) 
+ O\!\left( (m_t / m_{N_D})^4  \ln (m_{N_D}/m_t) \rule{0pt}{12pt} \right) ~~.
\ee
For $m_{N_D}$ near $m_t$ one needs to use 
Eq.~(\ref{fullNetb}). 
There is also a 2-body decay on $N_D$, proceeding through the small $W_L-W_R$ mixing:
\be 
\Gamma \left( N_D \to e^- W^+ \right) = \frac{g_R^4 \, M_W^2 \, \sin^2\! 2\beta}{ 64 \pi g^2 \, M_{W'}^4\! } \, m_{N_D}^3 
\left( 1 +2 \frac{M_W^2}{m_{N_D}^2} \right)    \left( 1 - \frac{M_W^2}{m_{N_D}^2} \right)^{\! 2}   ~~.
\label{eq:2body}
\ee

The $W'$ couplings shown in Eq.~(\ref{eq:leptons}) imply 
that the decays of $N_D$ involving $\tau$'s have opposite charges compared to the decays involving electrons \cite{Coloma:2015una}. 
The $N_D \to \tau^+ W^{\prime *}\!   \to \tau^+ \bar u d$ or $\tau^+ \bar c s$ widths are equal to that 
given in Eq.~(\ref{eq:3body}). Similarly, the $N_D \to \tau^+ W^{\prime *}\!   \to \tau^+ \bar t b$
and $N_D \!   \to \tau^+ W^-$ decays have the same widths as those shown in Eqs.~(\ref{eq:3bodytb}) and (\ref{eq:2body}), respectively.

\begin{figure}[t]
\centering
\includegraphics[width=0.7\textwidth]{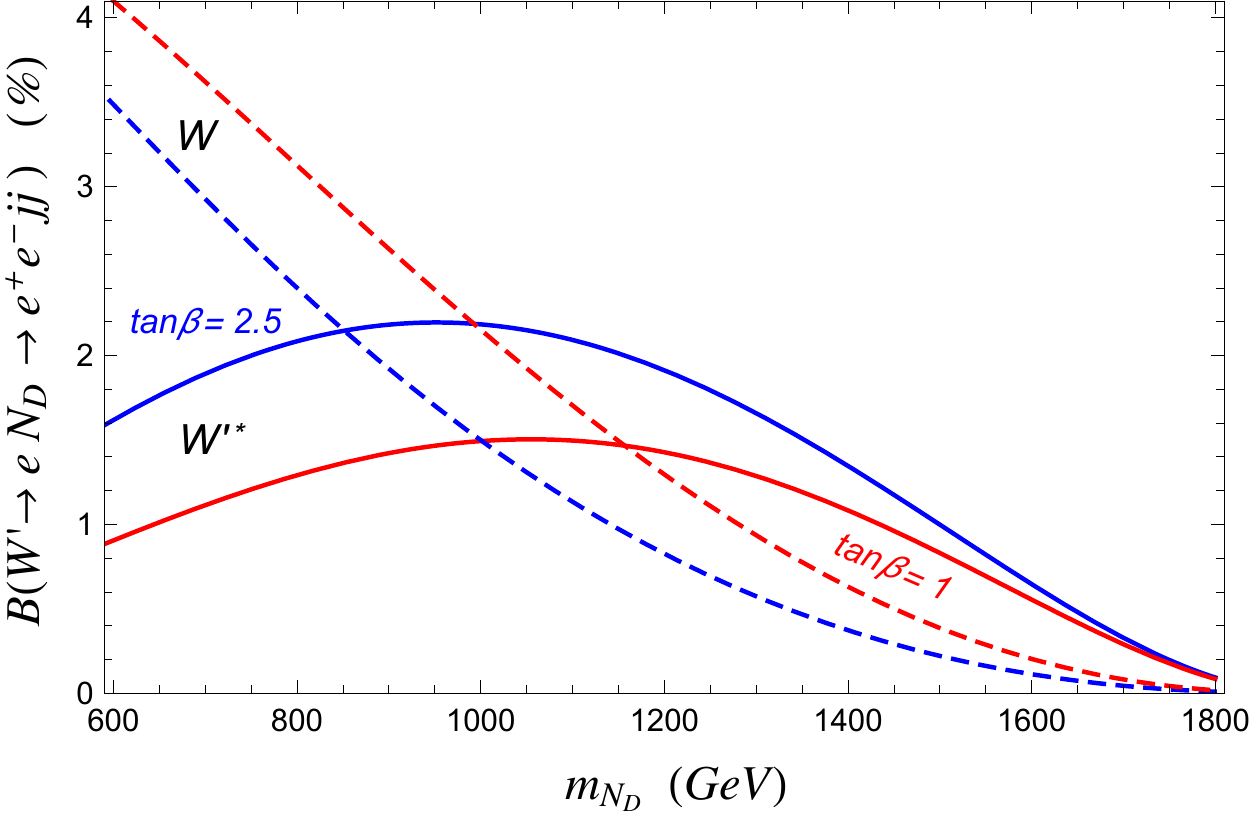}
\caption[]{Branching fractions of the cascade decays $W^\prime \!\to e N_D \to e^+e^-  W^{\prime *} \to e^+e^- jj$ (solid lines) through an off-shell $W'$,
and $W^\prime \!\to e N_D \to e^+e^- W \to e^+e^- jj$ (dashed lines) through a SM $W$,
where $j$ is a jet arising from a quark (or antiquark) of the first or second generation. 
The parameters used here are $\tan\beta = 1$ or 2.5, and $M_{W'}= 1.9$ TeV.
The Dirac fermion $N_D$ has couplings given in Eq.~(\ref{eq:leptons}).
}
\label{fig:Beejj}
\end{figure}

The branching fraction of the  $N_D$ 3-body decay into $ejj$ is
\be
B(N_D \to e^- W^{\prime *}\!   \to e^- j j ) =  \frac{1}{2} \left( 1 + \frac{1}{2}  f(m_{N_D})+ \frac{16\pi^2 M_W^2}{ g^2 \, m_{N_D}^2} \, \sin^2\! 2\beta \right)^{-1}   ~~,
\label{eq:BRejj}
\ee
where we ignored terms of order $M_W^4/m_{N_D}^4$. 
Multiplying this branching fraction by $B(W' \to eN_D)$ we obtain the total branching fraction shown in Figure~\ref{fig:Beejj}.
   
\begin{figure}[t]
\centering
\includegraphics[width=0.7\textwidth]{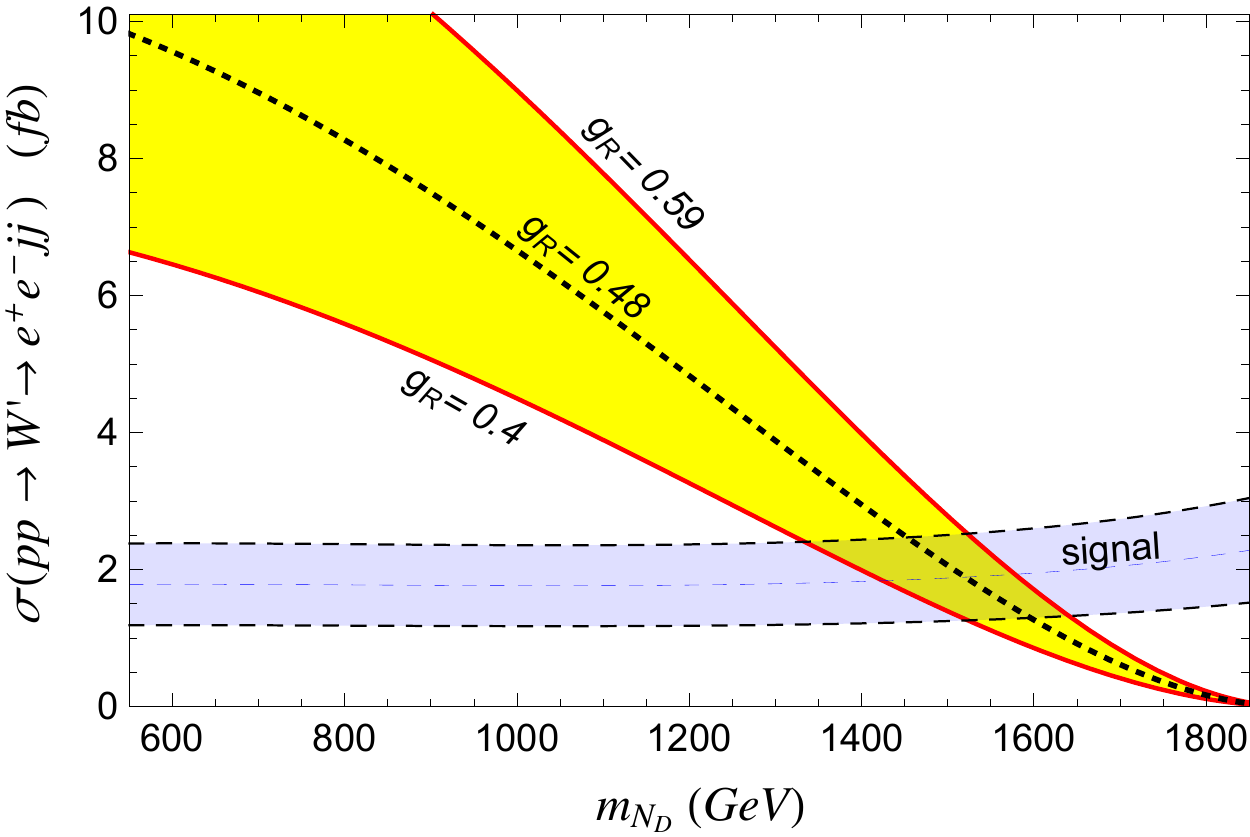} 
\caption[]{Cross section for $pp \to W^\prime \to e N_D \to e^+e^- jj$ at $\sqrt{s} = 8$ TeV for  $M_{W'}= 1.9$ TeV. 
The yellow shaded region between the two solid lines corresponds to the predicted cross section 
when  $g_R $ varies between 0.40 and 0.59.
The dotted line represents the preferred value $g_R =0.48$ (see Section 5.2).
The band between dashed lines (labelled ``signal") represents the cross section required to explain the CMS $e^+e^- jj$ signal ($10\pm 3.3$ events with $M_{eejj} > 1.8$ TeV).
}
\label{fig:sigmaeejj}
\end{figure}

Let us compare the CMS $e^+e^-jj$ signal \cite{Khachatryan:2014dka} with the predicted rate for $pp \to W^\prime \to e N_D \to e^+e^- jj$.
The acceptance-times-efficiency of the CMS event selection, estimated through 
our simulation, for events with $eejj$ mass above 1.8 TeV arising from the 3-body decay $N_D\to ejj$ 
decreases from 32\% to 26\% when $m_{N_D}$ grows from 0.5 TeV to 1.6 TeV. 
The analogous acceptance-times-efficiency ranges for $N_D\to et \bar b$  and  $N_D\to e^- W^+$  are 28\%--21\% and 8.4\%--7.2\%, respectively.
Given that there are 10 excess events,
and the luminosity is 19.7 fb$^{-1}$, we obtain a central value for the signal cross section of 0.84 fb. Including an uncertainty of $\pm 3.3$ events  
gives the horizontal shaded band labelled ``signal" in Figure \ref{fig:sigmaeejj}. 
The predicted cross section for $pp \to W^\prime \to e N_D \to e^+e^- jj$, given by
$\sigma_8 (W')  B(W^\prime \to e N_D \to e^+e^- jj)$, as a function of the $N_D$ mass is within the shaded region between 
the two solid lines in Figure \ref{fig:sigmaeejj}.
That shaded region corresponds to $\sigma_8 (W')$ in the 120--250 fb range ({\it i.e.}, $g_R=0.4$--0.59) 
and $\sin 2\beta =1$, but it is almost insensitive to  $\sin 2\beta$ in the 0.6--1 range ({\it i.e.}, $\tan \beta =1$--2.8). 
The large overlap of the predicted and observed signal regions is a significant success of our gauge-invariant $W'$ model.
The values of $m_{N_D}$ that fit the $eejj$ signal are between 1.35--1.65 TeV, with some dependence on $g_R$.

The dimension-5 operators (\ref{eq:ops}) arise from integrating out some fields. 
A simple choice is a set of three chiral fermions, $\psi_L^e$, $\psi_L^\mu$, $\psi_L^\tau$,  
which are $SU(3)_c \times SU(2)_L \times SU(2)_R \times U(1)_{B-L}$ singlets and carry global $U(1)$ charge $-1, 0, +1$, respectively. 
The most general mass terms invariant under the global $U(1)$ are $ m_\psi^{e\tau}  (\bar \psi_L^e)^c \, \psi_L^\tau + {\rm H.c.}$ and 
$ m_\psi^{\mu\mu}  (\bar \psi_L^\mu)^c \, \psi_L^\mu$. These chiral fermions have Yukawa couplings to 
the $SU(2)_R$-doublet leptons: 
\be
H_R \left( y_{e\psi} \overline L^e_R \psi_L^e + y_{e\psi} \overline L^\mu_R \psi_L^\mu + y_{\tau\psi} \overline L^\tau_R \psi_L^\tau \right)  + {\rm H.c.}
\ee 
Integrating out the $\psi_L$ fermions gives the operators (\ref{eq:ops}), with coefficients  $C_{e\tau} =  y_{e\psi} y_{\tau\psi} $ and $C_{\mu\mu} =  y_{\mu\psi}^2 $. 

\bigskip

\section{Bosonic $W'$ decays}\setcounter{equation}{0}

In addition to the couplings to SM particles (see Section~\ref{sec:Wpwidthandproduction})  and to right-handed neutrinos (see Section 3), 
the $W'$ also has couplings to the heavy Higgs bosons present in the $\Sigma$ bidoublet ($H^\pm, H^0, A^0$, as in Two-Higgs-doublet models) 
and in the $H_R$ doublet (a single neutral scalar, $h^0_R$).  
These couplings are due to the kinetic terms of the scalar fields and are determined by gauge invariance.  For simplicity we assume that the 
$\Sigma$ VEV is CP invariant, and that $h^0_R$ scalar is heavier than the $W'$ boson.

The  $W'$ coupling to the  $W$ and a heavy neutral Higgs boson arises (up to negligible corrections of order $M_W^2/M_{W'}^2$) 
from the kinetic term of $\Sigma$:
\be
\gR M_W W'^{+}_\mu W^{-\mu} \left( \cos (2\beta-\delta) \; H^0 +i \cos2\beta  \; A^0  \rule{0mm}{5mm} \right) + \mathrm{H.c.}
\ee
By contrast, the kinetic terms of $\Sigma$ and $H_R$ have comparable contributions to 
the $W'$ coupling to the $Z$ and the heavy charged Higgs boson, with the sum given by 
\be
-\gR M_Z \cos2\beta\;\, W'^{+}_\mu Z^\mu H^- + \mathrm{H.c.}~~.
\ee
The $W'$  couplings involving two charged Higgs bosons are, again,  almost entirely due to the $\Sigma$ kinetic term:
\be
\frac{\gR}{2} W'^{+}_\mu \; H^-\overleftrightarrow{\partial}^{\! \mu}  \left(\sin2\beta \; A^0 
- i \sin(2\beta-\delta) \;  H^0  
- i \cos(2\beta-\delta) \;  h^0
 \rule{0mm}{5mm}  \right) + \mathrm{H.c.}~~  
\ee

\begin{figure}[t] 
   \centering
   \includegraphics[width=0.75\textwidth]{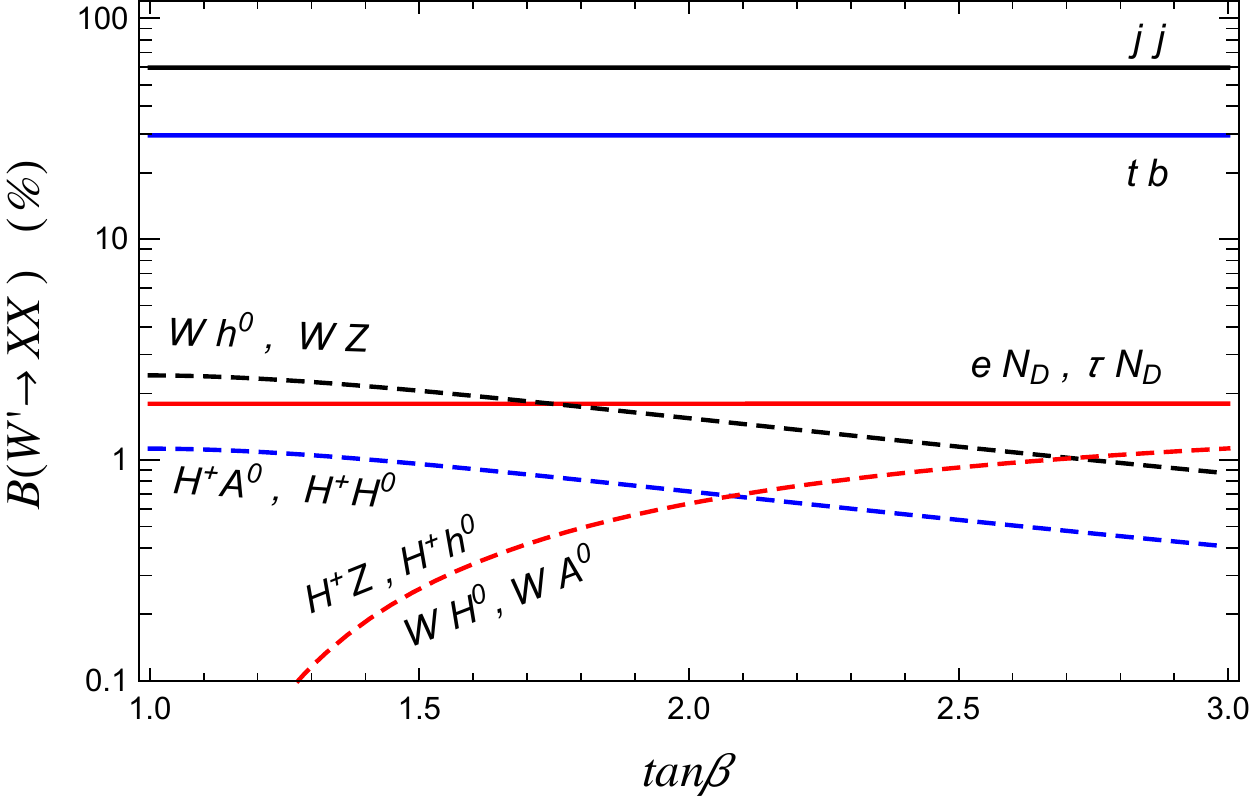} 
   \caption{Branching fractions of the $W'$ boson for $M_{W'} = 1.9$ TeV, $m_{N_D} =1.5$ TeV, and common mass of 0.6 TeV
   for heavy Higgs bosons, in the alignment limit ($\delta=0$). The lines labelled with two or more decay modes  give 
   the individual branching fractions. }
   \label{fig:WprimeBR}
\end{figure}

The $W'$  partial widths induced by the above couplings  are 
\bear
&& \Gamma(W'\rightarrow ZH^+) \simeq  \Gamma(W'\rightarrow WA^0)  \simeq   \frac{\gR^2\cos^2 2\beta}{192\pi}M_{W'}
\left( 1 - \frac{M_{H^+}^2 }{ M_{W'}^2 }  \right)^{\! 3}  ~,
\nonumber \\ [2mm]
&& \Gamma(W'\rightarrow WH^0)  \simeq  \Gamma(W'\rightarrow H^+h^0)  \simeq  \frac{\gR^2\cos^2 (2\beta-\delta)}{192\pi}M_{W'}
\left( 1 - \frac{M_{H^+}^2 }{ M_{W'}^2 }  \right)^{\! 3}   ~,
\nonumber \\ [2mm]
&& \Gamma(W'\rightarrow H^+A^0)  \simeq  \Gamma(W'\rightarrow H^+H^0) \simeq  \frac{\gR^2\sin^2 2\beta}{192\pi}M_{W'} 
\left( 1 - \frac{4 M_{H^+}^2 }{ M_{W'}^2 }  \right)^{\! 3/2}   ~,
\eear
where we have assumed that the heavy Higgs bosons have a common mass $M_{H^+}$ and are 
sufficiently light so that corrections of order $M_h^2/(M_{W'} - M_{H^+})^2$ can be ignored.  The corrections to each of these are given in \cite{Dobrescu:2015yba}.  As expected, the partial widths to modes related by Goldstone equivalence are equal, up to kinematic factors that have been ignored here.
The $W'$ branching fractions are displayed as a function of $\tan \beta$ in Figure \ref{fig:WprimeBR}.

The $3.4\sigma$ local excess in the ATLAS diboson resonance search with $WZ\rightarrow JJ$ \cite{Aad:2015owa} ($J$ is a merged jet arising from a boosted $W$ or $Z$)
corresponds to 13 observed events of invariant mass $1.85 \tev \le m_{JJ} \le 2.05 \tev$, with an expected background of 5 events.  In a related CMS analysis \cite{Khachatryan:2014hpa}, which does not attempt to distinguish between $W$ and $Z$ merged jets, there is a smaller $1.3\sigma$ excess.   
If the ATLAS diboson excess is due to a $W'$ decaying into  $WZ$ followed by hadronic decays of $W$ and $Z$, 
then excess events should eventually also appear in other $WZ$ decay channels at the same mass.  

The branching fraction for $W' \to WZ  \to \ell\ell\ell\nu$ is too small to be useful in Run 1 (we expect less than 1 event to be observed in 20 fb$^{-1}$ of 8 TeV data).
The semileptonic channels, $WZ \to \ell\nu J$ and $WZ \to J \ell^+\ell^-$, however, are as sensitive to a $W'$ as the $W' \to WZ\rightarrow JJ$ channel.
The CMS search in the  $WZ \to J \ell^+\ell^-$ channel has indeed yielded a $2\sigma$ excess at a mass in the 1.8--1.9 TeV range, while the 
CMS search in the  $WZ \to \ell\nu J$ channel, which is slightly more sensitive, has  only a $1\sigma$ excess in that mass range.
The ATLAS searches in the semileptonic channels, however, are in good agreement with the background estimate. Given that the sensitivity in all these channels is 
comparable, the ATLAS $WZ\rightarrow JJ$ result is in conflict at more than $2\sigma$  with the ATLAS combination of $WZ$ semileptonic and leptonic channels \cite{atlasdijetcombo}.
This discrepancy may be due to a statistical fluctuation. We point out, however, that the resolution of this issue may be more convoluted.
It may not be possible to analyze the $WZ$ channels separately from other $W'$ decay modes. 

Consider first the $W'\rightarrow jj$ channel. The hadronic jets produced by a light quark have invariant masses that may be  much larger 
than mass of the initial quark. Some of these jets may resemble a boosted $W$ or $Z$ decaying hadronically.  Through simulation, we estimate 
that approximately $1\%$ of the $jj$ events give two jets with jet masses close to the $W$ or $Z$ mass. 
Even if $\sim 40\%$ of these jets fail jet grooming requirements \cite{Aad:2015owa}, the branching fraction for $W' \to jj$ is $\sim 30$ larger than for
$W' \to WZ$ and so this mode contributes a significant number of events to the $JJ$ signal.  A precise simulation of the event selection and detector effects in this channel
is beyond the scope this work, but can be performed by the ATLAS and CMS collaborations. 

Similarly, contributions to what appear to be  $W'\rightarrow WZ\rightarrow JJ$ events may be due to other $W'$ signals, 
including $W'\rightarrow t b$, $W'\rightarrow W h^0 $, $W'\rightarrow H^+ A^0/H^0 \to (tb)(t\bar t)$, etc. It is tempting to speculate 
that the whole ATLAS diboson excess is due to these other signals, in which case there would be no lower limit on the $W' \to WZ$ branching fraction
 (or equivalently on $\sin^2\! 2 \beta$). However, the CMS  $WZ \to J \ell^+\ell^-$ excess seems to contradict this possibility.
 Furthermore, the CMS search \cite{CMS:2015gla} 
for a $Wh^0$ resonance, with $W\rightarrow \ell\nu$ and $h^0\rightarrow b\bar{b}$ has a $2.2\sigma$ excess, corresponding to 3 events with $Wh^0$ invariant mass between 1.8 TeV--1.9 TeV and  $0.3$ background events.  In the alignment limit, the $WZ$ and $Wh^0$ branching fractions are equal, so 
these observations place a lower limit on $\sin^2 2\beta$.
The $pp \to W'\rightarrow WZ$ cross section that would account for the $WZ \to J \ell^+\ell^-$ and $Wh^0 \to \ell\nu b\bar b$ events is roughly  in the 5--10 fb range,  assuming the selection acceptance-times-efficiency for $Wh^0$ to be similar to that of $WZ \to JJ$, 
$0.1\le A\times \epsilon \le 0.2$ \cite{Aad:2015owa}. 
This cross section indicates that these modes may have undergone a slight upward fluctuation and that $\sin 2\beta$ must be close to maximal.
Assuming that the entire $JJ$ excess does come from the $WZ$ final state places a similar constraint, 
$0.5 \lesssim \tan\beta \lesssim 2.0 $ for $\gR=0.4$, and $0.36 \lesssim \tan\beta \lesssim 2.8 $ for $\gR= 0.5$.

Recently, at Run 2 of the LHC, there have been searches for diboson final states.  CMS, using 2.6 fb$^{-1}$ of data, have searched for a resonance in both the $\ell\nu q\bar{q}$ and $JJ$ final states \cite{CMS:2015nmz}, while ATLAS, using 3.2 fb$^{-1}$ of data, have searched in the $JJ$ final state \cite{ATLASRun2WZJJ}.  Neither experiment has more than a $1\sigma$ excess in the region of $M_{W'}\approx 1.9$ TeV.  The CMS result places the strongest constraint: $g_R\lesssim 0.46$ at $M_{W'}=1.9$ TeV.  This bound is comparable to that coming from the Run 2 ATLAS dijet resonance search \cite{ATLAS:2015nsi}, which is sensitive to the $W'\rightarrow jj$ channel.  The $Wh$ resonance search at Run 2 \cite{ATLASWhRun2} has a small ($\sim 1.5\sigma$) excess at $M_{W'}=1.9$ TeV, consistent with the $Wh$ excess at the same mass in Run 1 \cite{CMS:2015gla}.

Another interesting diboson channel is $W' \to WZ$ followed by $Z \to \nu\bar \nu$ and $W\to jj$ \cite{Liew:2015osa}.
As the $W$ and $Z$ bosons are highly boosted, the first channel would appear as missing transverse energy ($\met$) and 
a merged jet ($J$) of mass near $M_W$ whose $p_T$ is back-to-back with the $\met$. The invariant mass distribution of the $J+\met$ system (``merged mono-jet")
has a plateau-like shape that extends between a few hundred GeV and $M_{W'}$. The mono-jet searches \cite{CMS:2015jha, Khachatryan:2014rra, Aad:2015zva} have produced 
an excess of events that may have a shape of this type, but it is too early to draw a conclusion as to whether its origin is new physics.  With an increase in data, this will be an interesting search channel both in merged mono-jet and using sub-structure techniques.

There are additional diboson channels that, with more accumulated luminosity, will provide nontrivial tests of the $W'$ properties.  These involve some of the smaller $W$, $Z$, or $h^0$ branching  fractions.  Final states involving three or more leptons receive contributions from multiple topologies, each with different kinematics,  {\it e.g.},  $W'\rightarrow WZ \rightarrow \ell\nu\ell\ell$ and $W'\rightarrow Wh^0 \rightarrow \ell\nu \ell\nu\ell\nu$.
Both $WZ$ and $Wh^0$ give contributions to final states with three taus. 

Besides decays to SM bosons, $W'$ bosons can generically decay into the heavy scalars associated with the 
Higgs sector responsible for breaking the extended gauge symmetry \cite{Dobrescu:2013gza}.
In the model discussed here, the $W'$ decays into heavy Higgs bosons  leads to top-rich final states, {\it e.g.}, $W'\rightarrow H^+ A^0\rightarrow t\bar{b}t\bar{t}$.  As was pointed out in \cite{Dobrescu:2015yba} these final states have the potential to explain the ATLAS same-sign leptons plus $b$-jets excess \cite{Aad:2015gdg}.   Due to differences in the structure of the right-handed neutrino sectors discussed here and in \cite{Dobrescu:2015yba}, the relative contributions from heavy Higgs and heavy neutrinos to the same-sign leptons plus $b$-jets final state is slightly altered.

\bigskip

\section{Properties of the $Z'$ boson} \setcounter{equation}{0}
\label{sec:Zpproperties}

In addition to the electrically-charged gauge bosons, $W_L^\pm$ and $W_R^\pm$  (whose linear combinations give the $W$ and $W'$ bosons),
the $SU(2)_L \times SU(2)_R\times U(1)_{B-L}$ gauge fields include three electrically-neutral  bosons: $W_L^{3 \mu}$, $W_R^{3 \mu}$, $A_{B-L}^\mu$. 
Two linear combinations of these acquire masses due to the scalar VEVs,
and become the $Z$ and $Z'$ bosons, while the third linear combination is the photon. 

The VEVs of the bidoublet $\Sigma$ and of the $SU(2)_R$ doublet  $H_R$ have the form
\be
\langle \Sigma \rangle = v_H
 \begin{pmatrix}
\cos\!\beta  & 0  \\
 0 &  e^{i\alpha_{_{\Sigma}}}  \sin\!\beta  \\
 \end{pmatrix}  \;\;\;    ~~,     \;\;\;    \;\;\;    
\langle H_R \rangle = 
 \begin{pmatrix}
0  \\
u_H 
 \end{pmatrix}     ~~,
 \label{eq:VEV}
\ee
where $v_H  \simeq  \sqrt{2} M_{W} / g \approx 174 $ GeV is the weak scale, and the $SU(2)_R\times U(1)_{B-L}$ breaking scale is
\be
u_H \simeq  \sqrt{2} \frac{ M_{W'} }{ g_R }  \approx 5\! - \!7 \; {\rm TeV} ~~, 
\ee
up to corrections of order $M_W^2/M_{W'}^2$.  
For simplicity, we assume that the CP-violating phase $\alpha_{_{\Sigma}}$ vanishes.
The VEVs in Eq.~(\ref{eq:VEV}) induce mass terms for the electrically-neutral gauge bosons,
\be
\frac{u_H^2}{4} \left( g_R  W_R^{3 \mu} - g_{B-L} A_{B-L}^\mu  \right)^2  + \frac{v_H^2}{4} \left( g_L  W_L^{3 \mu}  - g_R  W_R^{3 \mu} \right)^2   ~~,
\label{eq:Zpmassterms}
\ee
where $g_L \simeq g$ is the $SU(2)_L$ gauge coupling, and $g_{B-L}$ is the $U(1)_{B-L}$ gauge coupling, which is related 
to the SM hypercharge gauge coupling ($g_Y \approx 0.363$  at 2 TeV) by
\be
g_{B-L} = \left( \frac{1}{g_Y^2} -  \frac{1}{g_R^2} \right)^{\! -1/2}    ~~.
\label{eq:gBL}
\ee

The mass terms in Eq.~(\ref{eq:Zpmassterms}) lead to a $3\times 3$ mass matrix of rank 2, 
which is diagonalized by the following unitary transformation 
that relates the  photon ($A^\mu$) and the $Z$ and $Z'$ bosons to the $SU(2)_L \times SU(2)_R\times U(1)_{B-L}$ 
gauge fields:
\be
\left( \begin{array}{c}    W_{L}^{3 \mu}  \\ [4mm] W_{R}^{3 \mu}  \\ [4mm]  A_{B\!-\!L}^\mu \end{array} \right) =
 \left( \begin{array}{ccc}  
				 s_W              \; &  \;   c_W              \;  &  \;      -c_R \, {\displaystyle \frac{g_R M_W^2}{g_L M_{Z'}^2}  }\\ [5mm]
				 s_R  c_W    \;  &  \;   - s_R s_W    \;  &  \;     c_R  \\ [5mm]
			         c_R  c_W    \;  &  \;   - c_R s_W    \;  &  \;    -s_R 
    \end{array} \right)   
    \left( \begin{array}{c}    A^\mu  \\ [4mm] Z^\mu \\ [4mm]   Z^{\prime \mu} \end{array} \right)
    ~~,
\label{eq:zp}
\ee
where we kept only the leading order in $M_W^2/M_{Z'}^2$  for each element of the matrix. 
In analogy to the SM  $s_W  \equiv  \sin\theta_W$, we defined $s_R  \equiv  \sin\theta_R$ and $c_R  \equiv  \cos\theta_R$, with 
\be
c_R =  \frac{ g_R  }{ \sqrt{g_{B-L}^2+ g_R^2} }    ~~,
\ee
which implies a simple relation between $\theta_R$ and $g_R$:
\be
s_R = \frac{g_Y}{g_R} ~~.
\ee

The mass of the $Z'$ boson is related to the $W'$ mass by
\be
M_{Z'} = \frac{M_{W'}}{c_R} ~~.
\label{eq:Mratio}
\ee
As the ranges for  $M_{W'}$ and $g_R$ can be derived from the $W'$ signals, a range for $M_{Z'}$ can be predicted. 
We will show, however, that a more stringent  
lower limit on  $M_{Z'}$ in this model is imposed by  $Z'$ searches in Run 1 of the LHC.

\subsection{$Z'$ branching fractions}

The coupling of the $Z'$ boson to a chiral fermion $\psi$ is determined by the fermion's hypercharge ($Y$, in the normalization where 
the hypercharge of weak singlets equals their electric charge) 
and $SU(2)_R$ charge ($T^3_R$), and is given by
\be
\frac{\gR}{c_R} \, \left(T^3_R -Y s_R^2 \right)  \, Z'_\mu \bar{\psi} \gamma^\mu \psi ~~.
\label{eq:Zprimefermions}
\ee
As a result, the tree-level $Z'$ decay widths into lepton pairs are 
\bear
&& \Gamma (Z' \!\rightarrow e^+e^-)  = \Gamma (Z' \!\rightarrow \mu^+ \mu^-)   = \Gamma (Z' \!\rightarrow \tau^+\tau^-)   
=  \frac{g_R^2}{ 96 \pi c_R^2 }  \left( 1 - 4 s_R^2 + 5 s_R^4  \right)  M_{Z'}   ~,
\nonumber \\ [2mm]
&& \Gamma (Z' \!\rightarrow N_D \bar N_D)  = \frac{g_R^2}{ 48 \pi c_R^2 }  \,   M_{Z'}  
\left( 1 - \frac{4m_{N_D}^2 }{ M_{Z'}^2 }\right)^{\! 3/2}
 ~,
 \nonumber \\ [2mm]
&& \Gamma (Z' \!\rightarrow \nu \bar \nu)  = \frac{g_R^2}{ 32 \pi c_R^2 } \,  s_R^4 \,  M_{Z'}   ~.
\label{eq:ZpLeptons}
\eear
We assumed here that the $Z' \to N_R^\mu \bar N_R^\mu $  channel is kinematically closed.
The $Z'$ decay widths into quark-antiquark pairs are the same for the three fermion generations (the $m_t^2/M_{Z'}^2$ corrections are negligible), 
and given by
\bear
&& \Gamma (Z' \!\rightarrow u \bar u)  
=  \frac{g_R^2}{ 288 \pi c_R^2 }  \left( 9 - 24 s_R^2 + 17 s_R^4  \right)  M_{Z'}   ~,
\nonumber \\ [2mm]
&& \Gamma (Z' \!\rightarrow d \bar d)  = \frac{g_R^2}{ 288 \pi c_R^2 }  \,  \left( 9 - 12 s_R^2 + 5 s_R^4  \right)   M_{Z'}    ~.
 \eear

The $Z'$ also couples to pairs of gauge and Higgs bosons.  
The triple gauge boson couplings are fixed by gauge invariance, and the $Z'WW$ coupling is
\be
i \gR c_R\frac{M_W^2}{M_{Z'}^2} \left( \left(W^-_\mu Z'_\nu - W^-_\nu Z'_\mu\right)\partial^\mu W^{+\nu} +
\frac{1}{2}  \left(W^+_\mu W^-_\nu - W^+_\nu W^{-_\mu}\right)\partial^\mu Z'^{\nu}  \right)  +{\rm H.c.} ~~
\ee
The decay width is
\be
\Gamma (Z'\rightarrow W^+ W^-) =\frac{\gR^2 c_R^2}{192\pi} \, M_{Z'} ~~.
\label{eq:ZpWW}
\ee
If heavy enough, \ie\ $c_R<1/2$, the $Z'$ can decay to a pair of $W'$ bosons, 
\be
\Gamma(Z'\rightarrow W'W') = \frac{g_R^2}{192\pi c_R^2}   \left(1+20 \frac{M_{W'}^2}{M_{Z'}^2} +12 \frac{M_{W'}^4}{M_{Z'}^4} \right) \, \left(1-4 \frac{M_{W'}^2}{M_{Z'}^2} \right)^{\! 3/2} \, M_{Z'} ~~.
\ee

The $Z'$ has couplings to the five Higgs bosons coming from the $\Sigma$ bidoublet ($h^0, H^0, A^0$, $H^\pm$) as well as the single scalar ($h^0_R$) left after the doublet $H_R$ VEV breaks $SU(2)_R \times U(1)_{B-L}$ to $U(1)_Y$.  We will assume that $h^0_R$ is too heavy to be involved in $Z'$ decays.  
The relevant couplings controlling the decay of $Z'$ to final states involving two Higgs bosons are
\be
\frac{\gR c_R}{2}Z'_\mu \left(\sin \delta\; h \overleftrightarrow{\partial}^{\!\mu} A^0 +\cos \delta\; H \overleftrightarrow{\partial}^{\!\mu} A^0 + i H^- \overleftrightarrow{\partial}^{\!\mu} H^+\right)~.
\ee
If all the heavy Higgses are lighter than  $M_{Z'}/2$  the decay widths that survive in the alignment limit are
\bea
\Gamma(Z'  \!  \rightarrow H^0 A^0) &\simeq& \frac{\gR^2 c_R^2 \cos^2\!\delta}{192\pi} M_{Z'}\left(1-\frac{4M_{H^+}^2}{M_{Z'}^2}\right)^{\! 3/2}~,\nonumber\\ [2mm]
\Gamma(Z'  \!  \rightarrow H^+ H^-) &\simeq& \frac{\gR^2 c_R^2}{192\pi} M_{Z'}\left(1-\frac{4M_{H^+}^2}{M_{Z'}^2}\right)^{\! 3/2}~.
\label{eq:ZpHH}
\eea
where we assumed as in Section 4 that $M_{A^0} \simeq M_{H^0} \simeq M_{H^+}$.
The couplings of the $Z'$ to one Higgs boson and one gauge boson are
\be
\gR c_R M_Z  \, Z'_\mu Z^\mu \left( \cos\delta\; h^0  -\sin\delta\; H^0 \right)~,
\ee
and the corresponding width is
\be
\Gamma(Z' \! \rightarrow Z h^0) \simeq \frac{\gR^2 c_R^2 \cos^2\!\delta}{192\pi} M_{Z'}~.
\label{eq:ZpZh}
\ee
Note that the $Z$ boson here, as well as the $W$ bosons in Eq.~(\ref{eq:ZpWW}), are longitudinally polarized \cite{Hisano:2015gna}, so as expected based on 
the equivalence theorem the widths for $Z' \! \rightarrow WW$ and $Zh^0$ are approximately equal in the alignment limit. 
Furthermore, since the longitudinal $W$ and $Z$ are part of the same 
Higgs bidoublet field with the heavy Higgs bosons, the widths for $Z'  \!  \rightarrow H^0 A^0$ and $H^+ H^-$ shown in  Eq.~(\ref{eq:ZpHH}) are equal 
to those for $Z' \! \rightarrow Zh^0$ and $WW$ in the limit where phase space suppression is negligible.

There are two further modes that are negligible in the alignment limit,
\be
\Gamma(Z' \! \rightarrow Z H^0) \simeq \Gamma(Z' \!\rightarrow h^0 A^0) \simeq 
\frac{\gR^2 c_R^2 \sin^2\!\delta}{192\pi} M_{Z'}\left(1-\frac{M_{H^+}^2}{M_{Z'}^2}\right)^3 ~.
\ee
The decays to $H^+ W^-$, $H^+ W'^-$ and $W W'$ are suppressed by $M_Z^2/M_{Z'}^2$, and so are too small to be of interest at present. 
There are also 3-body decays that have small branching fractions and we ignore here.

\begin{figure}[t] 
\vspace*{-0.3cm}
   \centering
   \includegraphics[width=0.75\textwidth]{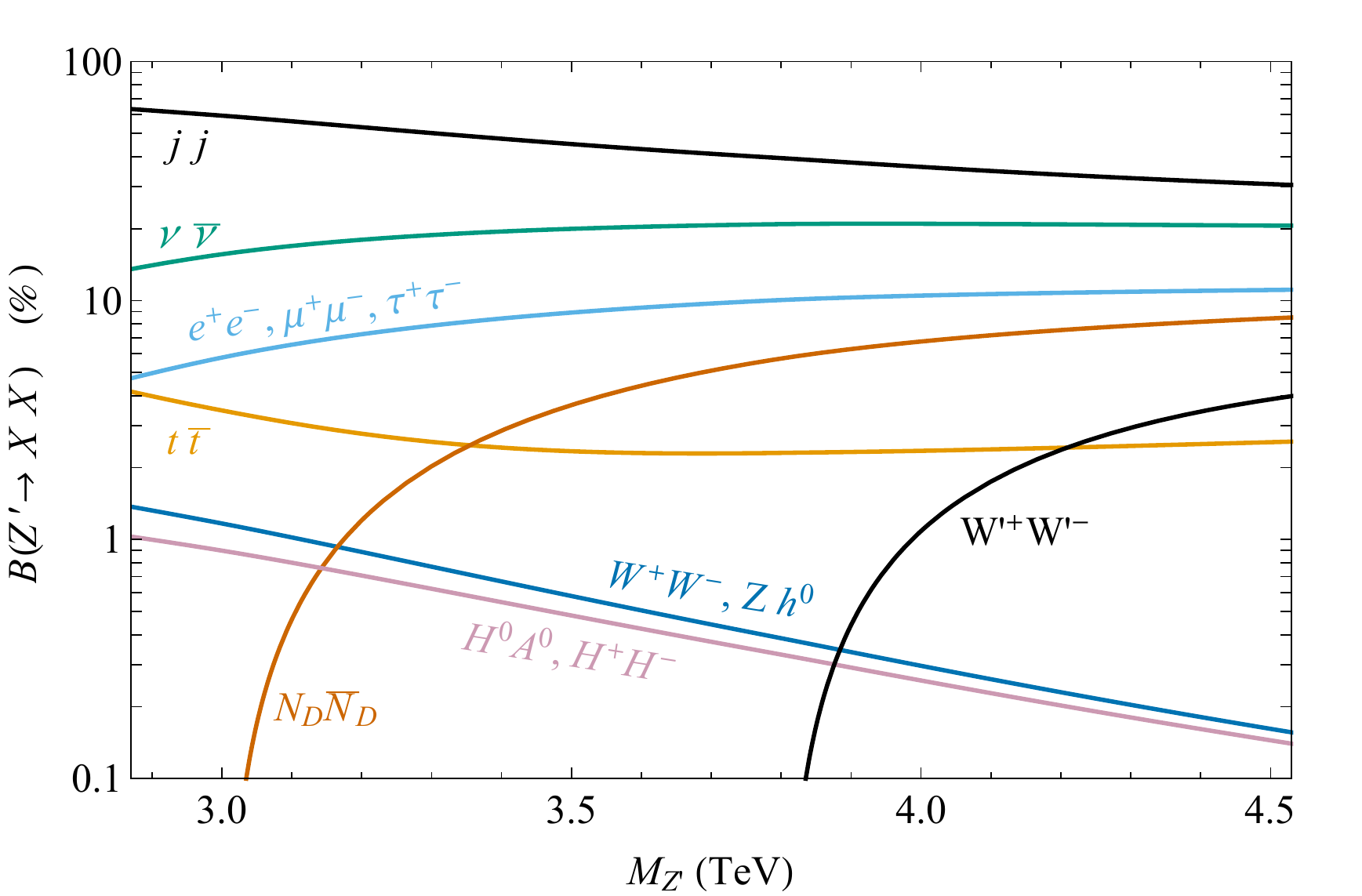} 
   \caption{Branching fractions of the $Z'$ boson for $M_{W'}=1.9$ TeV, $m_{N_D} =1.5$ TeV, and 
   $M_{H^+} = M_{H^0} = M_{A^0} = 600$ GeV 
 in the alignment limit ($\delta=0$).  The $jj$ line is the sum of the partial widths for $u\bar{u},\, d\bar{d},\, s\bar{s},\,c\bar{c}$, and $b\bar{b}$; the $\nu\bar{\nu}$ line is the sum of partial widths into SM neutrinos;  the lines labelled with two or more decay modes give the individual branching fractions.
}
   \label{fig:ZprimeBR}
\end{figure}

Based on Eqs.~(\ref{eq:ZpLeptons})-(\ref{eq:ZpZh}), we find the $Z'$ branching  fractions shown in 
 Figure~\ref{fig:ZprimeBR}. 
 The parameters used there are $M_{W'}=1.9$ TeV (other values of $M_{W'}$ in the 1.8--2 TeV range imply a different $s_R$, 
 leading to slight changes in the $Z'$ branching fractions), $m_{N_D}=1.5$ TeV (the branching fractions other than $N_D \bar N_D$ are rather insensitive to $m_{N_D}$),
 and a common mass of 600 GeV for the heavy Higgs bosons from the bidoublet (only the $H^0 A^0$ and $H^+ H^-$ branching fractions are sensitive to this choice). 
 Note that in the alignment limit the partial widths have no dependence on $\tan\beta$.  
 One should keep in mind that if the assumptions $m_{N^\mu} > M_{Z'}/2$ and $M_{h_R} > M_{Z'}$ used here turn out to be false, the 
 $Z' \to N_R^\mu \bar N_R^\mu $ and  $Z' \to Z h_R^0$ channels may prove to be important.

\bigskip

\subsection{$Z'$ signals at the LHC }

From Eq.~(\ref{eq:Zprimefermions}) it follows that the dominant $Z'$ production at hadron colliders is through its coupling to right-handed down quarks. 
The $Z'$ couplings depend on $g_R$, which in turn depends on $M_{Z'}$ for fixed $M_{W'}$ [see Eq.~(\ref{eq:Mratio})].
For $M_{W'} = 1.9$ TeV and $g_R$ in the range shown in Eq.~(\ref{eq:gR}) we find $M_{Z'}$ in the 2.4--4.5 TeV range.\footnote{Similar $M_{Z'}$ ranges have been discussed in related models \cite{Collins:2015wua}.} 
As no dilepton events of invariant mass 
above 1.8 TeV has been observed in Run 1 of the LHC, the lower part of the $M_{Z'}$ mass range can be ruled out. 
To derive the lower limit on  $M_{Z'}$, we require less than three predicted $Z' \to \ell^+\ell^-$ events in the dilepton searches
performed by ATLAS \cite{Aad:2014cka} and CMS \cite{Khachatryan:2014fba} taken together.
The $Z'$ production cross section at the 8 TeV LHC, obtained by multiplying the MadGraph result 
with $K(Z') = 1.16$  \cite{Aad:2014cka},  decreases from 1.6 fb for $M_{Z'} = 2.8$ TeV to 1.1 fb for $M_{Z'} = 2.9$ TeV. 
The acceptance-times-efficiency in the dielectron (dimuon) channel is $65\% \ (40\%)$ at ATLAS \cite{Aad:2014cka}
and $55\% \ (80\%)$ at CMS \cite{Khachatryan:2014fba}. Using the $Z' \to \ell^+\ell^-$ branching fraction shown in Figure \ref{fig:ZprimeBR},
we conclude that the lower limit on  $M_{Z'} $ set by the Run 1 searches is approximately 2.85 TeV.
Thus, the range of  $M_{Z'} $ relevant for this model is 2.9 TeV $ \lesssim M_{Z'}  \lesssim 4.5$ TeV.

\begin{figure}[t] 
\vspace*{-0.5cm}
   \centering
   \includegraphics[width=0.66\textwidth]{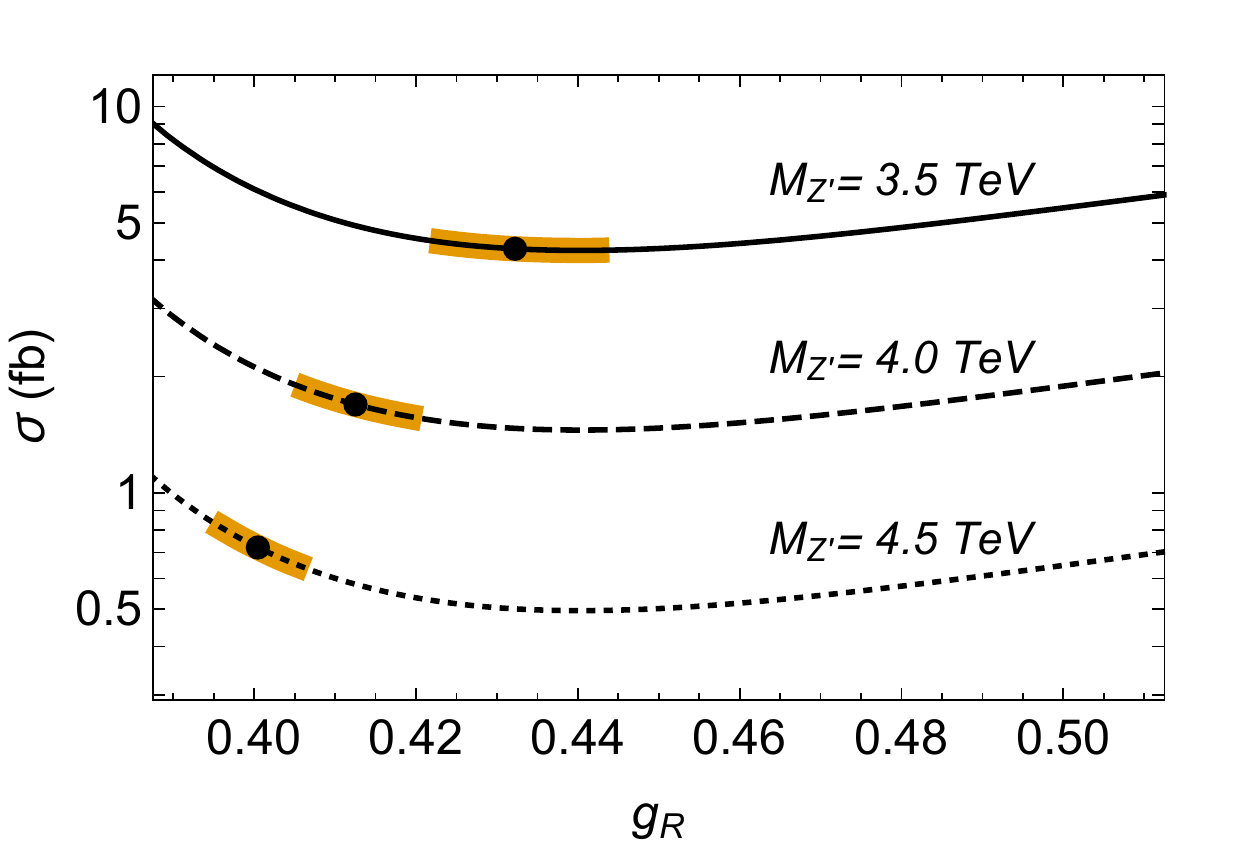} 
   \caption{Production cross section, including a $K$-factor of 1.16, for a $Z'$ boson of mass 3.5, 4, and 4.5 TeV 
   at the LHC with $\sqrt{s}=13$ TeV.  The thicker region of each curve denotes the range of $g_R$ for which $1.8 \le  M_{W'} \le 2$ TeV, 
   with the marked point at $M_{W'} =1.9$ TeV.}
   \label{fig:Zprimeproduction}
\end{figure}

The $Z'$ production cross section at the 13 TeV LHC, computed at leading order with MadGraph and multiplied by $K(Z') = 1.16$ 
is shown in Figure~\ref{fig:Zprimeproduction}.  A $Z'$ whose mass saturates the Run 1 bound has a production cross section of $\sim 19$ fb at Run 2, which predicts approximately five dilepton $Z'$ events after 5 fb$^{-1}$.  Although after 65 pb$^{-1}$ CMS observed a dielectron event \cite{CMSBigEvent} with invariant mass of $2.9$ TeV (the probablity for this event to be due to the SM background is $\sim 10^{-3}$) the more recent Run 2 analyses of CMS \cite{CMS:2015nhc} and ATLAS \cite{ATLASdileptonRun2} have no deviation from the SM prediction.  The ATLAS result, using 3.2 fb$^{-1}$, places a lower bound of 3.4 TeV on the $Z'$ mass, corresponding to $g_R=0.44$ and a total width of $\Gamma_{Z'} \approx 80$ GeV.

The dominant branching fraction is $Z' \!\to jj$ (see Figure 4). Furthermore,  
Higgs bosons lighter than $\sim 1$ TeV and SM bosons or top quarks, produced in the decay of the $Z'$, will be sufficiently boosted that their decay products will lie in a single jet.  Thus, the effective dijet branching fraction for a 2.9 TeV $Z'$, and $\gR=0.48$, increases from $62\%$ to about $70\%$, with the remaining $30\%$ due to $5\%$ for each of $e^+e^-$, $\mu^+\mu^-$ and $\tau^+\tau^-$, and a $14\%$ invisible branching fraction; we have assumed that $N_D$ is heavier than $M_{Z'}/2$.  Assuming the acceptance-times-efficiency for these high $p_T$ jets coming from $Z'$ decay is approximately $50\%$ at both CMS and ATLAS we estimate that the signal in 
each experiment is  $\sim 8$ dijet events with invariant mass around $3$ TeV in Run 1.  This signal is too small compared to the  QCD background 
\cite{Khachatryan:2015sja, Aad:2014aqa}.

After 30 fb$^{-1}$ of Run 2 each experiment, assuming a similar acceptance and efficiency as in Run 1, should have $\sim 200$ dijet events.  
The sizable QCD background makes this a challenging search channel, although it may be possible to use various signal features, including angular distributions, 
quark-versus-gluon discriminants, and  
substructure techniques (for the hadronic decays of boosted objects) to reduce the background.

\bigskip

\section{Conclusions}
\label{sec:conclusions}

The CMS and ATLAS Collaborations, analyzing LHC Run 1 data, have accumulated an intriguing set of excess events over disparate final states.  
These include  excesses near 2 TeV  in $jj$, $WZ$, $Wh$ and $e^+e^-jj$ resonance searches \cite{Khachatryan:2014dka}-\cite{Aad:2015owa}, 
as well as an excess in the final state with same-sign leptons plus $b$ jets \cite{Aad:2015gdg}. 
 We have shown that 
there is a common explanation for all these excess events in a model with $ SU(3)_c\times SU(2)_L\times SU(2)_R\times U(1)_{B-L}$ gauge structure, 
a Higgs sector with only a bidoublet and an  $SU(2)_R$ doublet,  and a flavor symmetry that controls the masses of the  right-handed neutrinos. 
We refer to this as the R$e\tau D$ model, to emphasize  that there is a flavor symmetry acting on the $e$ and $\tau$ 
right-handed neutrinos, and that the $SU(2)_R\times U(1)_{B-L}$ symmetry is broken by a Doublet. 

Let us briefly mention some differences between this model and the ``RDi$T$ model" presented in \cite{Dobrescu:2015qna,Dobrescu:2015yba} and the  ``R$e\tau T$ model" 
presented in \cite{Coloma:2015una}.
Both RDi$T$ and  R$e\tau T$ use an $SU(2)_R$ Triplet to break the $SU(2)_R\times U(1)_{B-L}$ gauge symmetry. As a result, the relation between 
the $W'$ and $Z'$ masses shown in Eq.~(\ref{eq:Mratio}) for our R$e\tau$D model is modified by a factor of $\sqrt{2}$,  pushing the $Z'$ mass
above 3.5 TeV. The RDi$T$ model has Dirac masses for right-handed neutrinos, with the left-handed Dirac partners provided by vectorlike 
$SU(2)_R$-doublet leptons. As a result, there is an additional parameter that controls the $W'$ coupling to an electron and a Dirac right-handed neutrino.
By contrast, in the  R$e\tau T$ model and our $Re\tau D$ model the $e$ and $\tau$ right-handed neutrinos form a Dirac fermion ($N_D$), whose mass
must be in the 1.4--1.6 TeV range (see Figure 2) to explain the CMS $e^+e^-jj$ excess. 

The excess events  from Run 1, taken together, point towards a $W'$ of mass $\sim 1.9$ TeV, which means that the decay products of the $W'$ are highly boosted.  The $jj$ excess is not only sensitive to $W'$ decays to light quarks but also to its decays to heavy SM particles, which subsequently decay hadronically.  This raises the \emph{effective} hadronic branching fraction from $\sim 60\%$ to $80-90\%$.  We take this effect into account and reanalyze the dijet excess and determine that the range necessary to explain the excess is $0.4\lesssim \gR \lesssim 0.59$, corresponding to a $W'$ width of $20-48$ GeV.  We also point out that the lack of an excess in the CMS $tb$ resonance search places a nontrivial constraint, $\gR\lesssim 0.45$, while the constraints from ATLAS are considerably weaker and allow the whole coupling range favored by the dijet excess.

An upper limit on $g_R$ is placed by dilepton resonance searches in Run 2: the $M_{Z'} \gtrsim 3.4$ TeV limit corresponds to $g_R \lesssim 0.44$ in this R$e\tau D$ model.  A comparable limit on $g_R$ is set by the Run 2 dijet searches.
The only other relevant parameters are $\tan\beta$, which controls the $W'\to WZ$ and $W'\to Wh^0$ branching fractions, and the masses of the heavy Higgs bosons. 
The Run 1 diboson events 
indicate $0.35\lesssim \tan \beta \lesssim 2.8$ (see Section 4), while the ATLAS events with same-sign leptons plus $b$ jets prefer $M_{H^+} \simeq M_{H^0} \simeq M_{A^0} \approx 500$--600 GeV \cite{Dobrescu:2015yba}. 

 Our MadGraph model files \cite{ourMG5} include all the new particles discussed here: $W'$, $Z'$, $N_D$, $H^+$, $H^0$, $A^0$. These allow simulations 
 of the many signals predicted in this model, and can be used by the experimental collaborations to disentangle the cross-contamination of new-physics channels 
 ({\it e.g.}, $W' \to jj$ versus $W' \to WZ \to JJ$, mentioned in Section 4).

There are many $W'$ decays that can be tested in Run 2. From Figure 3, it follows that there are at least 7 channels with branching fractions at percent level or above.
The $N_D$ fermion produced in two of these channels has several decay 
modes: $e jj$, $e tb$, $eW$ (as well as others with small branching fractions, including $eWh$ and $e WZ$) and the same with $e$ replaced by $\tau$. We thus
obtain more than 17 final states (without even counting the various decay modes of the SM bosons) in which the LHC Run 2 will have 
sensitivity to the existence of the $W'$ boson.

The $Z'$ presents a separate set of opportunities. Taking into account the $W'$ mass and production cross section that fit the Run 1 dijet excess and the lower $M_{Z'}$ limit set by Run 2, the $Z'$ mass is in the  $3.4-4.5$ TeV range.  The $Z'$ branching fractions, shown in Figure 4, are large enough to allow sensitivity in 9 channels,
with many additional ones (originating from $N_D \bar N_D$ and $W'W'$) opening up for large enough $M_{Z'}$.


\bigskip\bigskip

{\it \bf Acknowledgments:} We would like to thank Dan Amidei, David Calvet, John Campbell, Richard Cavanaugh,  Pilar Coloma, Frank~Deppisch, Wade Fisher,  
Mike Hance, Philip Harris, Robert Harris, James Hirschauer, Zhen Liu, 
Jeremy Love, Sho Maruyama, Steve Mrenna, Natsumi Nagata, Jianming Qian, Martin Schmaltz,  Jacob Searcy, Karishma Sekhon, Nhan Tran, 
Bing Zhou and Junjie Zhu,  for helpful conversations and comments. 

\bigskip\bigskip

%
\renewcommand{\theequation}{A.\arabic{equation}}
\section*{Appendix: \ Right-handed neutrino width }
\label{appendix}\setcounter{equation}{0}

The 3-body decays of the right-handed neutrino (see Section 3) proceed through an off-shell $W'$.
The squared matrix element for the $N_D \to e t \bar b$ decay, summed over final states and averaged over the initial one is
\be
\frac{3g_R^4 }{ 2 \left( m_{12}^2 \! - \! M_{W'}^2\right)^2} \!
\left[ m_{23}^2  \left( \!
m_{N_D}^2 \!\! +m_t^2 \! - m_{23}^2 \!
- \frac{m_{N_D}^2 m_t^2}{M_{W'}^2} \right)
+ \frac{m_{N_D}^2 m_t^2}{4M_{W'}^4} \left(m_{N_D}^2 \!\! - m_{12}^2\right) \left(m_{12}^2 \! - m_t^2\right) \right] ~.
\ee
Here $m_{12}$ and $m_{13}$ are the invariant masses of the $t \bar b$ and $eb$ systems, respectively.
After integrating over phase space we find the following width:
\bear    \hspace*{-1.7cm} 
&& \hspace*{-.7cm} \Gamma\! \left( N_D \! \to 
e^- t \, \bar b  \, \right)  =
\frac{3 g_R^4  \, m_{N_D}^5}{ \! (8\pi)^3 M_{W'}^4 \!}  \!
\left[   \, \frac{M_{W'}^6}{m_{N_D}^6} \! 
\left( \! 1 \! - \frac{m_t^2}{ m_{N_D}^2} \right)\!   \left( \! 1\! -\frac{m_{N_D}^4\! +m_t^4 - m_{N_D}^2 m_t^2/2  \! }{6 M_{W'}^4}  + \frac{m_{N_D}^4 m_t^4 }{4M_{W'}^8}  \right)  \right.
\nonumber \\ [2mm]
&& + \, \frac{M_{W'}^8}{m_{N_D}^8} \! \left( \!1 \!- \frac{m_{N_D}^2}{M_{W'}^2} \right) \!
     \left( \!1\!- \frac{m_t^2}{M_{W'}^2} \! \right)  \!
     \left( \!1\!  - \frac{m_{N_D}^2 m_t^2}{4M_{W'}^4} \! \left( \!1\! + \!\frac{m_t^2}{M_{W'}^2}\right) \!\! \left( \!1\! +\!\frac{m_{N_D}^2}{M_{W'}^2} \right) \!\right) \ln \frac{M_{W'}^2 \! -m_{N_D}^2}{M_{W'}^2 \! -m_t^2}\!
\nonumber \\ [2mm]
&&  \left. - \, \frac{M_{W'}^4}{2 m_{N_D}^4} \left( 1 - \frac{m_t^4}{m_{N_D}^4} \right) \left( 1 + \frac{5 m_{N_D}^2 m_t^2 }{4 M_{W'}^4}   \right) +
2   \frac{m_t^4}{m_{N_D}^4}  \left(1  + \frac{m_{N_D}^2 m_t^2}{4M_{W'}^4} \right) \ln \frac{m_{N_D}}{m_t}  
\right]  ~~.
\label{fullNetb}
\eear
In the limit $m_t \to 0$ this is equal to the width for $N_D \to e^- u \bar d$:
\be  \hspace*{-0.4cm} 
\Gamma \left( N_D \to e^- u \bar d \right) = 
\frac{3 g_R^4 M_{W'}^4}{ \! (8\pi)^3 \, m_{N_D}^3 \!}  \! \left[
\left(  \!  1  \! -\frac{m_{N_D}^2}{M_{W'}^2}\right) \ln \left( \!  1  \! -\frac{m_{N_D}^2}{M_{W'}^2} \right)+
\frac{m_{N_D}^2}{M_{W'}^2} - \frac{m_{N_D}^4}{2M_{W'}^4} - \frac{m_{N_D}^6}{6M_{W'}^6}
\right] ~.
\ee
Expanding this expression in $m_{N_D}^2/M_{W'}^2$ gives Eq.~(\ref{eq:3body}), while 
expanding Eq.~(\ref{fullNetb})  in both $m_t/m_{N_D}$ and $m_{N_D}/M_{W'}$ gives Eq.~(\ref{eq:3bodytb}).

%

\vfil
\end{document}